\documentclass[twocolumn,aps,prb]{revtex4}
\usepackage{graphicx}
\usepackage{amssymb}
\usepackage{amsmath}
\usepackage{bm}
\usepackage{color}

\def\Journal #1,#2,#3,#4#5#6#7{#1 {\bf #2}, #3 (#4#5#6#7)}

\def\e{\varepsilon}
\def\p{\prime}
\def\s{\sigma}

\def\bk{\bm{k}}

\begin{document}

\title{Magnetotransport in the Weyl semimetal in the quantum limit \\
--- the role of the topological surface states}
\author{Yuya Ominato and Mikito Koshino}
\affiliation{Department of Physics, Tohoku University, Sendai 980-8578, Japan}
\date{\today}

\begin{abstract}
We theoretically study the magnetoconductivity of the Weyl semimetal having a surface boundary
under  ${\bf E}||{\bf B}$ geometry
and demonstrate that the topological surface state plays an essential role in the magnetotransport.
In the long-range disorder limit where the scattering between the two Weyl nodes vanishes,
the conductivity diverges in the bulk model (i.e., periodic boundary condition)
as usually expected, since the direct inter-node relaxation is absent.
In the presence of the surface, however,
the inter-node relaxation always takes place through the mediation by the surface states,
and that prevents the conductivity divergence.
The magnetic-field dependence becomes also quite different between the two cases,
where the conductivity linearly increases in $B$ in the surface boundary case,
in contrast to $B$-independent behavior in the bulk-periodic case.
This is an interesting example in which 
the same system exhibits completely different properties
in the surface boundary condition and the periodic boundary condition
even in the macroscopic size limit.
In the short-range regime where the direct intervalley scattering is dominant,   
the surface states are irrelevant and the conductivity approaches that of the bulk periodic model.
\end{abstract}
\maketitle

\section{Introduction}
The Weyl semimetal is a novel type of 3D material characterized by the zero-gap band structure 
and the topological surface states. 
There the energy bands are touching at more than two isolated points in the momentum space,
and these band-touching points (Weyl nodes) are connected by the surface-state bands.
\cite{murakami2007phase, burkov2011weyl, burkov2011topological, wan2011topological, yang2011quantum,hosur2013recent}
Recently the enormous effort has been made in searching for the Weyl semimetal phase in the real materials,
and the evidences of the surface states were observed in several experiments. 
\cite{xu2015observation,weng2015weyl,lu2015experimental,xu2015discovery,lv2015experimental,huang2015weyl,xu2015experimental}

In the Weyl semimetals, it has long been known that 
application of magnetic field ${\bf B}$ and electric field ${\bf E}$ in a  parallel fashion
gives rise to an unusual transport property, which is closely related to the chiral anomaly. 
\cite{nielsen1983adler,aji2012adler,son2013chiral,burkov2014chiral,gorbar2014chiral,parameswaran2014probing,lu2015high,burkov2015negative}
There the electrons in one valley ($k$-space region around a Weyl node) 
are pumped to another valley in an irreversible fashion,
causing an imbalance in the number of electrons at the different valleys.
Such a population imbalance leads to an electric current in parallel to ${\bf E}\parallel{\bf B}$,
and this is expected to be observed as a negative magnetoresistance.
\cite{nielsen1983adler,aji2012adler,son2013chiral,burkov2014chiral,gorbar2014chiral,lu2015high,burkov2015negative}
In the weak-field regime, the theoretical analysis predicts the magnetoresistance proportional to $B^2$.
\cite{son2013chiral,burkov2014chiral,burkov2015negative}
Experimental signatures of the negative magnetoresistance were also reported.
\cite{kim2013dirac,li2016negative,yang2015chiral,zhang2015detection,li2015giant,zhang2015observation,PhysRevX.5.031023,xiong2015evidence}
The previous theoretical work on the negative magnetoresistance, however,
assumed the bulk system without the boundary, and it is not clear
how such the property is influenced by the topological surface state,
which is another distinctive feature of the Weyl semimetal.

In this paper, we study the magnetoconductivity of the Weyl semimetal having a surface boundary
in the ${\bf E}\parallel{\bf B}$ geometry,
and demonstrate that the topological surface state plays an essential role in this problem.
Here we consider a tight-binding lattice model of the Weyl semimetal having a pair of band-touching points with a surface boundary.
When a magnetic field is applied in parallel to the surface, 
we show that the surface band is seamlessly connected to the 
zero-th Landau level of the bulk spectrum, and form a closed Fermi surface.
\cite{PhysRevB.93.081103}
Here we consider the high-field limit (quantum limit) where
only the zero-th Landau level is relevant for the electronic transport.
The quantum limit is achieved in condition that
the Landau level spacing exceeds the other energy scales such as
the thermal broadening and the level broadening caused by impurity scattering.

For this situation, we calculate the conductivity
under the electric field ${\bf E}$ parallel to ${\bf B}$ using the Boltzmann transport theory
in the presence of the scattering potentials.
For the disorder potential, 
we introduce the Gaussian-type potential 
which effectively models various scatterers
from the short-range scatters such as lattice defects to the long-range scatters
such as screened charged impurities, by changing the spacial range of the potential.
In the long-range disorder regime where the scattering between the Weyl nodes is negligible,
the inter-node relaxation is banned and then 
the conductivity diverges in the bulk model with the periodic boundary condition, as usually expected.
In the presence of the surface, however,
the inter-node relaxation always takes place through the mediation by the surface states, 
and that prevents the divergence of the conductivity.
The magnetic-field dependence becomes also quite different between the two cases,
where the conductivity linearly increases in $B$ in the surface boundary model,
while it is independent of $B$ in the bulk model.
The discrepancy between the periodic and surface boundary conditions
remains prominent even in the macroscopic size limit.
In the short-range limit where the direct intervalley scattering is dominant,
on the other hand, the surface states becomes irrelevant and 
the conductivity approximates that for the bulk periodic system.
It should be mentioned that the interference of the surface states and bulk states
was studied for a different geometry with the magnetic field 
perpendicular to the surface, \cite{potter2014quantum,zhang2015quantum,baum2015current}
while here we consider the parallel geometry.
The effect of the topological surface states in the Weyl semimetal
was also studied in terms of anomalous Hall conductivity  \cite{haldane2014attachment}
and cyclotron resonant phenomena.\cite{baum2015current}

The paper is organized as follows.
In Sec.\ \ref{sec_model}, we introduce a lattice model
which represents a Weyl semimetal with a pair of Weyl nodes.
In Sec.\ \ref{sec_Boltz}, 
we present the formalism to calculate the conductivity using the Boltzmann transport theory.
In Sec.\ \ref{sec_infinite}, we argue about the conductivity with the periodic boundary condition,
and in Sec.\ \ref{sec_surface}, we show the numerical results and present an analytical argument
for the conductivity with the surface boundary condition.
The discussion and conclusion are given in Sec.\ \ref{sec_dis}.

\section{Tight-binding model}
\label{sec_model}

\begin{figure}
\begin{center}
\leavevmode\includegraphics[width=1.\hsize]{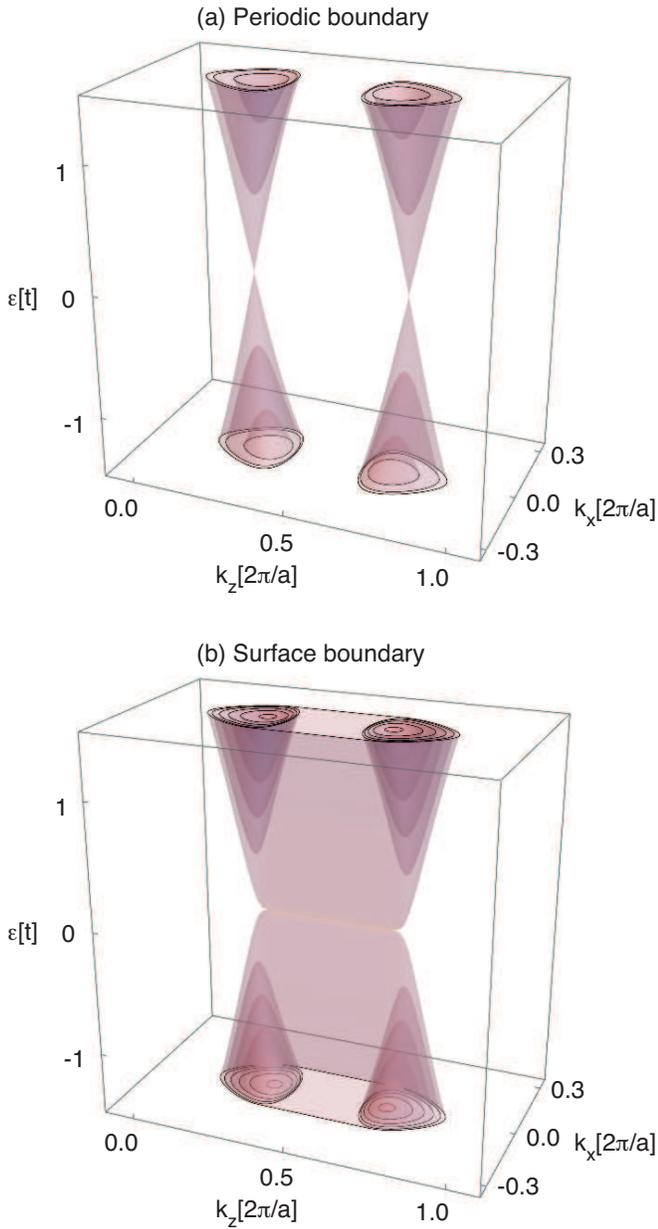}
\end{center}
\caption{Band structure in zero magnetic field
calculated for $L_y = 20 a$ with (a) periodic and (b) surface boundary conditions.
}
\label{fig_band_0}
\end{figure}

We consider a tight-binding lattice model 
expressed by the Hamiltonian,\cite{yang2011quantum,PhysRevB.93.081103}
\begin{align}
H &= 2 t (\sigma_x\sin k_x a + \sigma_y\sin k_y a + \sigma_z \cos k_z a)   \nonumber \\
    & \qquad + 2 m \sigma_z (2-\cos k_x a-\cos k_y a),
\end{align}
where $\sigma_i\,(i=x,y,z)$ are the Pauli matrices, $a$ is a lattice constant.
In $|m| > |t|/2$, the conduction and valence bands are touching only at two Weyl nodes at $(0,0,\pm\pi/(2a))$,
and the low-energy band structure around the nodes is almost unaffected by the parameter $m$.
In the following calculation we set $m=t$ for simplicity.
The existence of the boundary surfaces leads to a significant change to the spectrum of this system.
Figure \ref{fig_band_0} compares the band structures with and without the surface:
we consider a finite system size
with $L_y = 20 a$ in $y$-direction,
and impose the periodic boundary condition in (a)  and the surface boundary condition in (b).
In both cases we assume the system is infinite in $x$ and $z$ directions and
plot the energy bands against $(k_x, k_z)$. 
In the surface boundary systems (b), we see that the surface band
connects the bulk bands at the two Weyl nodes. 

Now we introduce an external magnetic field $B$ parallel to $z$ axis (i.e. parallel to the surface)
by taking the Landau gauge  $\bm{A}=(-By,0,0)$.
The eigenstate is written as $e^{i(k_x x+k_z z)}\psi (y)$ with 
the Bloch wave vectors $\bm{k}=(k_x,k_z)$ 
and 2-component spinor $\psi$.
The Schr\"odinger equation for $\psi (y)$ becomes
\begin{align}
(\varepsilon/t) \psi(y) &= (-i\s_x-\s_z)e^{ik_x a} e^{i 2\pi \phi y /a}\psi(y) \notag \\ 
&  + (i\s_x-\s_z)e^{-ik_x a}e^{-i 2\pi \phi y /a}\psi(y) \notag \\ 
&  + (i\s_y-\s_z)\psi(y+a)+(-i\s_y-\s_z)\psi(y-a) \notag \\ 
&  + \s_z(4+e^{ik_z a}+e^{-ik_z a})\psi(y),
\label{sch_eq_lattice}
\end{align}
where $\varepsilon$ is the eigenenergy and $\phi=B a^2/(h/e)$ is the magnetic flux penetrating a unit cell.

\begin{figure}
\begin{center}
\leavevmode\includegraphics[width=0.95\hsize]{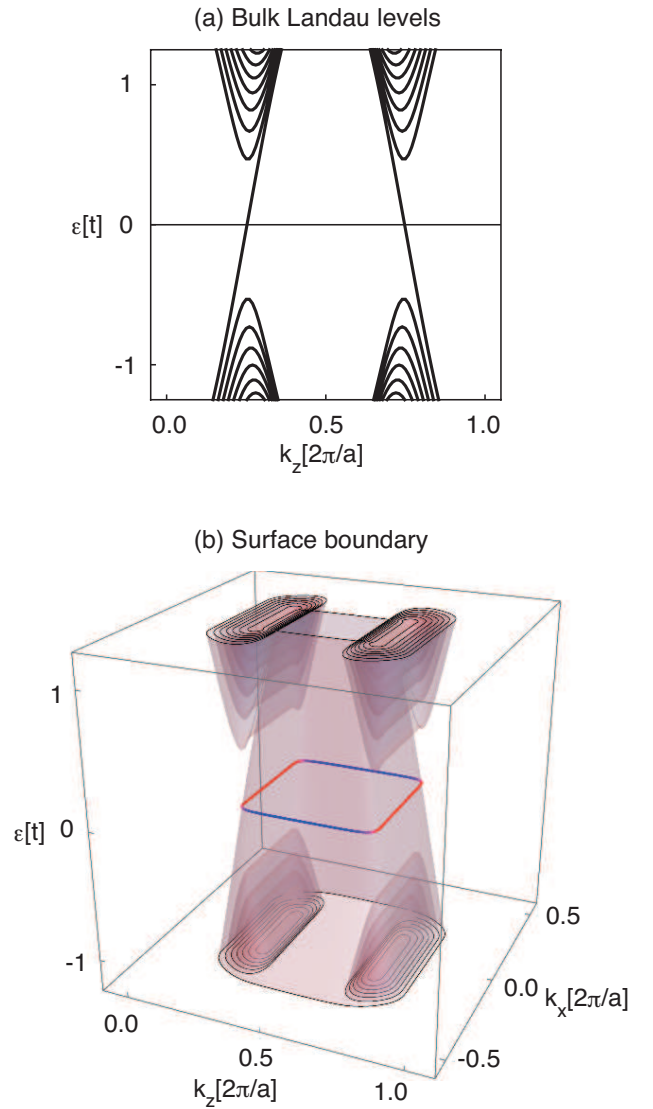}
\end{center}
\caption{Band structure in a finite magnetic field $\phi=0.005$,
calculated for (a) periodic boundary condition and
(b) $L_y = 100a$ with surface boundary condition.
}
\label{fig_band_1}
\end{figure}

\begin{figure}
\begin{center}
\leavevmode\includegraphics[width=1.\hsize]{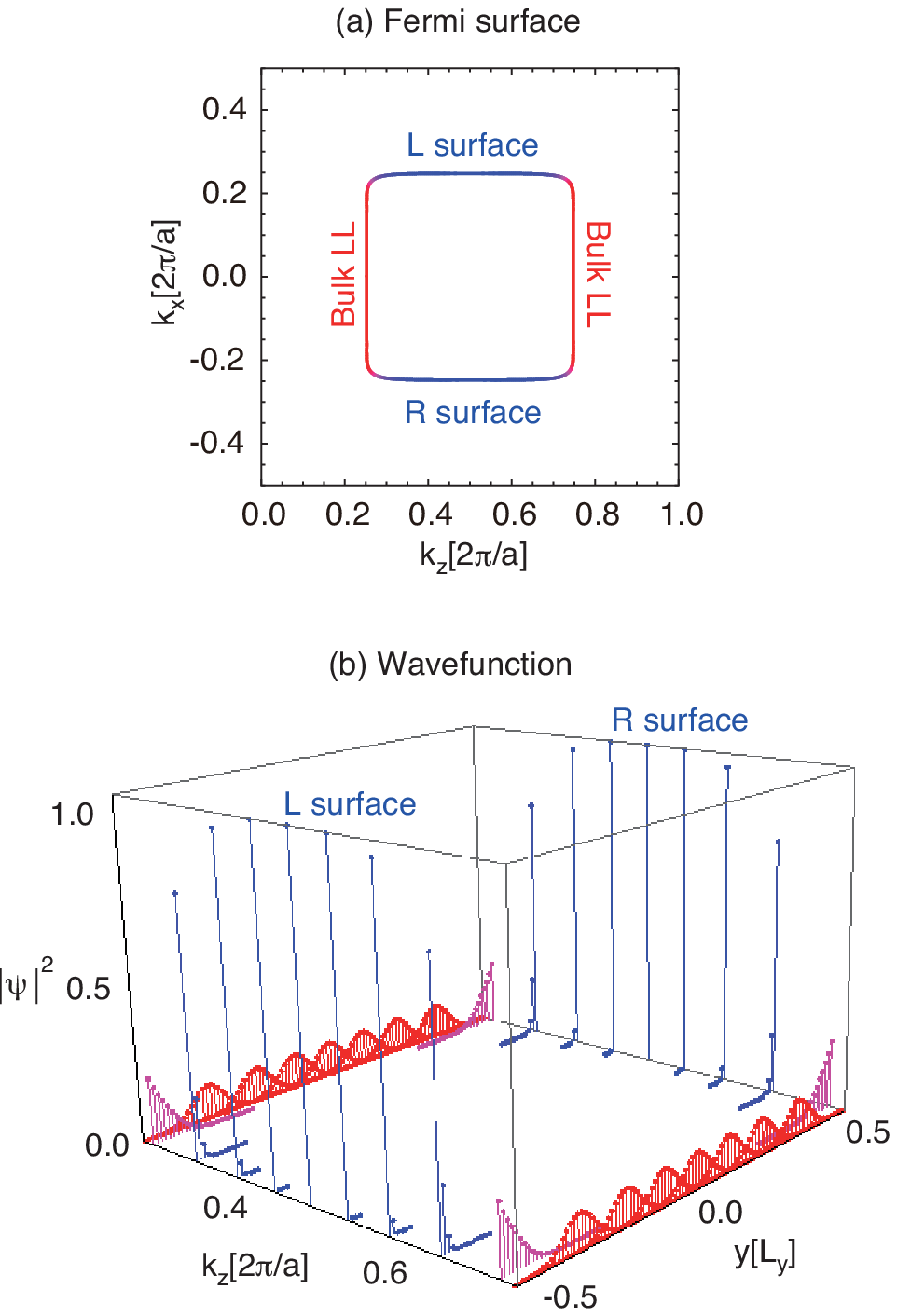}
\end{center}
\caption{(a) Fermi surface and
(b) the wavefuctions in $y$-direction for several states on the Fermi surface.
We set $L_y = 100a$ and $\phi= 0.005$.}
\label{fig_fs}
\end{figure}

Figure \ref{fig_band_1} is the similar figure under a finite magnetic field $\phi= 0.005$.
In the periodic condition (a),
the Landau levels are only dispersing in $k_z$ direction.
The Fermi surface at $\e=0$ is composed of
a pair of straight lines crossing the zero-th Landau levels of two valleys.
In the surface boundary system (b), on the other hand,
the Fermi surface  becomes a rectangular-like shape,
where the two sides along $k_x$ are mainly contributed by the zero-th Landau levels,
while the other two sides along $k_z$ are by the surface states.
The width of the Fermi surface in $k_x$
is $L_y/l_B^2$ exactly corresponding to the bulk Landau level degeneracy
where $l_B=\sqrt{\hbar/(eB)}$ is the magnetic length,
and the width in $k_z$ direction is $\pi/a$. 
Figure \ref{fig_fs} shows (a) the Fermi surface and
(b) wavefuction in $y$ direction plotted separately for different $k_z$'s,
where we set $L_y = 100a$ and $\phi= 0.005$.


\section{Boltzmann transport theory}
\label{sec_Boltz}

For the obtained Fermi surface at $\e=0$, 
we calculate the conductivity under the electric field ${\bf E}$ parallel to $z$ direction
using the Boltzmann transport theory.
The Boltzmann theory is valid as long as the Fermi surface is well defined,
i.e., the Fermi surface is almost unchanged within the range of the energy broadening.
In zero magnetic field, the approximation fails at zero Fermi energy where the Fermi surface becomes a point,
but in the quantum limit (high $B$-field), we always have a finite-sized Fermi surface 
as shown in Fig. \ref{fig_band_1}(b), and then the Boltzmann treatment becomes valid
under a moderate energy broadening not to mix other subbands.
Boltzmann transport equation is written in usual manner as
\begin{align}
\frac{\partial n_{\bk}}{\partial t}=\frac{e}{\hbar}\bm{E}\cdot\frac{\partial n_{\bk}}{\partial \bk}
                   				    +\left(\frac{\partial n_{\bk}}{\partial t}\right)_{\rm coll},
\label{boltzmann_eq_chap3}
\end{align}
where
\begin{align}
\left(\frac{\partial n_{\bk}}{\partial t}\right)_{\rm coll}&=\sum_{\bk^\p}
										   (n_{\bk^\p}-n_{\bk})W_{\bk\bk^\p},
										   \notag \\
W_{\bk\bk^\p}&=\frac{2\pi}{\hbar}\langle|V_{\bk\bk^\p}|^2\rangle\delta(\e_{\bk}-\e_{\bk^\p}),
\end{align}
and $n_{\bk}$ is the distribution function,
and $\langle\cdots\rangle$ represents the average over the configuration of the impurity positions.
We define $\delta n_{\bk}$ as $n_{\bk}=n_{\bk}^0+\delta n_{\bk}$ with $n_{\bk}^0$ is the Fermi distribution function.
We define $h_{\bk}$ and a relaxation time $\tau_{\bk}$ as
\begin{align}
\delta n_{\bk}&=(-e)E_z\tau_{\bk} h_{\bk}\left(-\frac{\partial n^0(\e)}{\partial \e}\right)_{\e=\e_{\bk}}
\label{eq_gk}, \\
\frac{1}{\tau_{\bk}}&=\sum_{\bk^\p}W_{\bk\bk^\p},
\label{def_tau} 
\end{align}
and then the Boltzmann equation becomes
\begin{align}
h_{\bk}=v_{\bk z}+\sum_{\bk^\p}\tau_{\bk^\p}h_{\bk^\p}W_{\bk\bk^\p},
\end{align}
where $v_{\bk z} = (\partial \e_{\bk}/ \partial k_z)/\hbar$.
Solving this self-consistent equation, we obtain $h_{\bk}$, and it gives $\delta n_{\bk}$.
The electric current density is derived by $j_z=(-e/V)\sum_{\bk}v_{\bk z}\delta n_{\bk}$ with the system volume $V$,
and the conductivity
at the zero temperature
$\s_{zz} = j_z/E_z$ is finally obtained as
\begin{align}
\sigma_{zz}=\frac{e^2}{L_y}\frac{1}{(2\pi)^2}\int {\rm d}\bk v_{\bk z}\tau_{\bk}h_{\bk}\delta(\e_{\rm F}-\e_{\bk}).
\label{cond_zz}
\end{align}

For the disorder potential, 
we introduce the Gaussian potential 
\begin{align}
V(\bm{r})&=\sum_i \frac{u_0}{C^3} \exp\left(-\frac{|\bm{r}-\bm{r}_i|^2}{d_0^2}\right), 
\end{align}
where $\bm{r}_i$ is the impurity position in the lattice
and $C$ is the normalization constant defined by
$C=\sum_{n=-\infty}^\infty a\exp\left[-(na/d_0)^2\right]$
so that the summation of a single impurity potential over all the lattice points becomes $u_0/a^3$.
We define the volume density of the scatterers as $n_{\rm i}$.
In the following, we define the long (short) range regime 
as a situation where  $d_0$ is much longer (shorter) than 
the inverse of the separation of Weyl nodes, $\pi/a$.
As we see below, the results are significantly different between the two regimes.

The Gaussian potential effectively models various type of scatterers
from lattice defects to charged impurities, by changing the potential range $d_0$.
If we consider charged impurities, for example, the Fourier transform of a single impurity
is given by $u_{\rm c}(q)=(4\pi e^2/\kappa)(q^2+q_{\rm s}^2)^{-1}$ with
the inverse screening length $q_{\rm s}$ and the static dielectric constant $\kappa$.
Noting that the Fourier transform of Gaussian potential is $u_{\rm G}(q)=u_0 \exp(-q^2d_0^2/4)$,
we can effectively simulate the charged impurity case by replacing $u_0$ with $4\pi e^2/(\kappa q_{\rm s}^2)$,
and $d_0$ with $1/q_{\rm s}$ in the Gaussian result.
This scheme does not cover bare Coulomb impurity $(q_{\rm s}=0)$ which has power-law decaying,
but here we consider the quantum limit with relatively large magnetic field, 
and there it is expected that the nonzero density of states of the Landau level
gives a finite electronic screening length.


\section{Conductivity of the infinite system}
\label{sec_infinite}

First, we calculate the bulk conductivity, i.e., the conductivity of the periodic system without the surface.
The bulk energy band in magnetic field becomes completely independent in $k_x$,
and then the Fermi surface becomes just two parallel lines at $k_z = \pm \pi/(2a)$ along $k_x$ direction.
In this case, we can analytically solve the Boltzmann transport equation 
and the conductivity is explicitly written as
\begin{align}
\sigma_{zz}^{\rm bulk}(d_0)=\sigma_0\exp\left[\frac{\pi^2}{2}\frac{d_0^2}{a^2}\right].
\label{cond_zz_bulk}
\end{align}
Here $\sigma_0$ is defined by 
\begin{align}
\sigma_0=\frac{1}{2\pi}\frac{e^2\hbar v^2}{n_{\rm i}u_0^2},
\end{align}
which characterizes the conductivity,
where $v=2ta/\hbar$.
The detail of the derivation of Eq.\ (\ref{cond_zz_bulk}) is given in the Appendix.
The only assumption to obtain the analytic formula Eq.\ (\ref{cond_zz_bulk})
is $l_B \gg d_0$, while it is not required in the numerics below.
The conductivity exponentially increases 
with increase of the potential range $d_0$.
This is because the two bulk branches at $k_z = \pm \pi/(2a)$,
traveling in the positive and negative directions, respectively,
cannot be relaxed when the scattering potential is too smooth. 
This property, i.e., a diverging conductivity in the independent valley limit, 
was predicted in the early study and it is regarded as an important consequence
of the chiral anomaly. \cite{nielsen1983adler}
We also notice that the conductivity does not depend on the magnetic field $B$. 
Here the number of states (i.e., the Landau level degeneracy)
increases as $\propto B$, but at the same time 
the relaxation time drops in inversely proportional to 
the number of states, resulting in the $B$-independent conductivity.\cite{aji2012adler}

\begin{figure}
\begin{center}
\leavevmode\includegraphics[width=0.8\hsize]{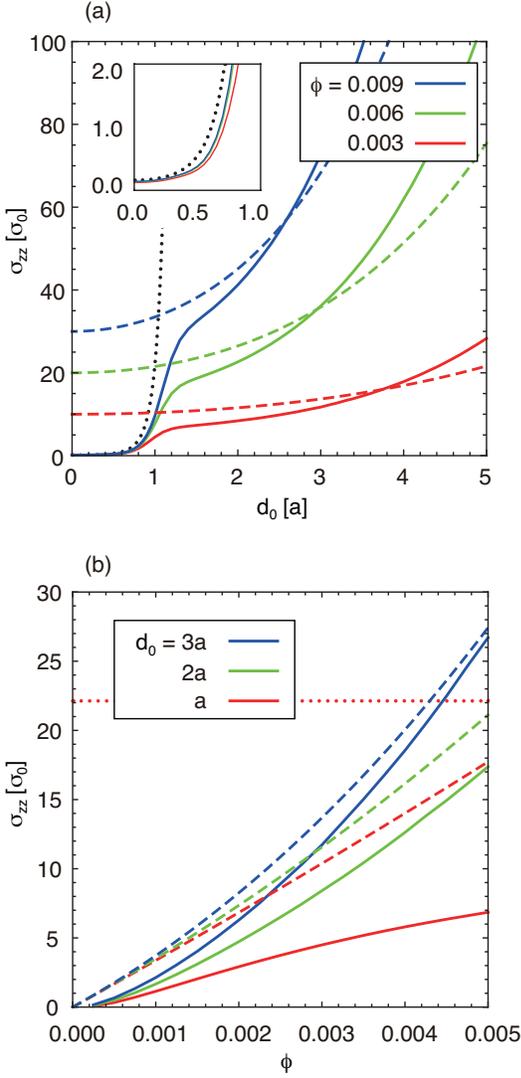}
\end{center}
\caption{(a) Conductivity as a function of the potential range $d_0$, 
at different magnetic fields $\phi=0.003$, 0.006, and 0.009. $L_y$ is set to $100a$.
The solid curve indicates the numerically obtained conductivity,
dotted black curve is the analytic expression for bulk (periodic) case, Eq.\ (\ref{cond_zz_bulk}),
and dashed curve is that for the surface-boundary case, Eq.\ (\ref{cond_zz_edge}).
The inset shows the expanded plot in the small $d_0$ region.
(b) Similar plot as a function of the magnetic field $\phi$ at $d_0=a, 2a, 3a$.
Red dotted curve is the bulk conductivity  Eq.\ (\ref{cond_zz_bulk}) for $d_0=a$.
}
\label{fig_cond}
\end{figure}

\begin{figure}
\begin{center}
\leavevmode\includegraphics[width=1\hsize]{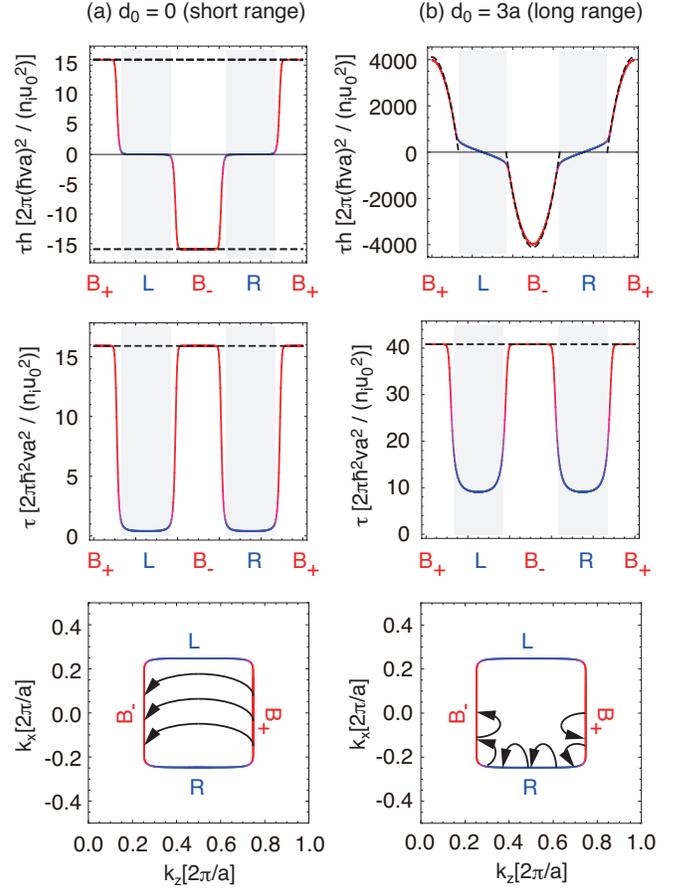}
\end{center}
\caption{
Plots of $\tau_{\bk}h_{\bk}$ ($\propto \delta n_{\bk}$; upper panel) 
and $\tau_{\bk}$ (middle panel) as functions of the $k$-space coordinate along the Fermi surface,
for (a) short-range $(d_0=0)$ and (b) long-range $(d_0=3a)$ cases
at the magnetic field $\phi=0.005$.
$L_y$ is set to $100a$.
The dashed curves indicate the analytic expressions
in the Appendix.
Lower panels: Schematic picture of the dominant relaxation path.
}
\label{fig_nk}
\end{figure}

\section{Conductivity with surface boundary}
\label{sec_surface}

The feature of the magnetotransport
greatly changes in the presence of the surface.
Here we consider the surface boundary system 
with the finite thickness of $L_y = 100a$ in $y$-direction, 
and numerically solve the Boltzmann equation (\ref{cond_zz})
by discretizing the Fermi surface
(a closed line in Fig. \ref{fig_band_1}(b))
into 400 $k$-points.
Fig.\ \ref{fig_cond}(a) shows the numerically calculated conductivity as a function of $d_0$, 
at several different magnetic fields $\phi=0.003$, 0.006, and 0.009.
The inset shows the expanded plot in the small $d_0$ (short-range) region.
The dotted black curve is the analytic bulk conductivity, Eq.\ (\ref{cond_zz_bulk}).
The numerical result fits well to
the bulk conductivity in the short-range region near $d_0 =0$. 
When increasing $d_0$, however,
the numerical solution deviates from the exponential increase, 
but crosses over to different, power-law behavior.


To understand the underlying physics, 
we show $\delta n_{\bk}$ on the Fermi surface 
for the short-range case  ($d_0 = 0$)
and the long-range case  ($d_0 = 3a$) in  Fig.\ \ref{fig_nk}(a) and (b), respectively.
In each coloumn, the top and middle panels
plot $\tau_{\bk}h_{\bk} \propto \delta n_{\bk}$
and $\tau_{\bk}$, respectively,
where the horizontal axis is the $k$-space coordinate along the Fermi surface.
The lowest panel shows the Fermi surface and schematic scattering processes by arrows.
In the short-range case (a),  the excessive carriers from the equilibrium 
are distributed only in the bulk Landau level
region. It is almost flat within each of the positive- and negative-velocity branches,
and suddenly jumps to zero when entering the surface region.
This reflects that the two bulk branches are well coupled by the direct scatterings,
and the situation is rather similar to the bulk.
In the long-range case (b), on the other hand, the distribution of the bulk states 
gradually changes along the whole Fermi surface.
There the two bulk branches are not coupled by the direct impurity scattering,
but the internode relaxation takes place through hopping via the surface states,
as schematically shown in the lowest panel.
This prevents the exponential increase of the conductivity in the long-range regime.


Actually we can explain these qualitative behavior
by an analytic approximate solution of the Boltzmann equation as shown in the following.
Here we consider the surface boundary system in the long range disorder region,
and completely neglect the direct scattering between the two branches of the bulk Landau levels.
We also assume that $\delta n_{\bk}$ is non-zero only in the bulk states
while it completely vanishes in the surface states.
This is equivalent to the assumption that the momentum relaxation is much faster
in the surface states than in the bulk states,
and
it is approximately satisfied in the numerical calculation in middle panel of Fig.\ \ref{fig_nk}(b).
The Boltzmann equation can be solved analytically and we obtain the conductivity
\begin{align}
\sigma_{zz} &=\frac{1}{3}\sigma_0
                           \left(1+\frac{d_0^2}{l_B^2}\right)^2\frac{L_y^2}{l_B^2}.                      
\label{cond_zz_edge}
\end{align}
As in the bulk case, we again assumed $l_B \gg d_0$ to obtain these  formulas
and the detail of the derivation is given in the Appendix.
The biggest difference from the bulk case Eq.\ (\ref{cond_zz_bulk})
is that the conductivity is a polynomial in $d_0$
and does not exponentially increase in the long-range limit $d_0\gg a$.
Second, the conductivity is now dependent on the magnetic field $B(\propto 1/l_B^2)$. 
In the low-field regime $l_B \gg d_0$,
it increases linearly to $B$, while it becomes super-linear in higher fields.
Lastly, the conductivity is dependent on the thickness and proportional to $L_y^2$.
Unlike conventional metals, the electrons in the present system
are scattered backward via the surface states, 
and this surface-mediated relaxation gives rise to unusual system-size dependence of the conductivity.

In Fig.\ \ref{fig_nk}, the approximate analytic expressions for $\tau_{\bk} h_{\bk}$
and  $\tau_{\bk}$ 
[see Appendix]
are indicated by the dashed lines in the top and middle panel, respectively.
The conductivity $\sigma_{zz}$ of Eq.\ (\ref{cond_zz_edge}) is also plotted 
dashed curves in Fig.\ \ref{fig_cond}(a).
We see that the approximate formula qualitatively explain the numerical curves
in the long range regime $d_0 = 3a$.
The most obvious discrepancy is that
$\tau_{\bk} h_{\bk}(\propto \delta n_{\bk})$ of the numerical curve is not actually vanishing in the surface state region,
and this is due to the finite relaxation time $\tau_{\bk}$ in the surface states
which is neglected in the analytics.
We can show that the surface distribution becomes more relevant in increasing $d_0$,
and this accounts for the deviation of the numerical and analytical 
conductivity in Fig.\ \ref{fig_cond}(a) in large $d_0$.

The crossover from the bulk periodic limit Eq.\ (\ref{cond_zz_bulk}) to the 
surface boundary regime Eq.\ (\ref{cond_zz_edge}) 
takes place when the two magnitudes are comparable.
Since the latter is proportional to $L_y^2$, the critical potential range $d_0$ of the crossover
depends on the system size $L_y$.
However, as the bulk conductivity increases very rapidly as  $\propto \exp(d_0^2/a^2)$,
the crossover point always come to $d_0/a = O(1)$ unless $L_y$ is exponentially large.
Therefore the huge discrepancy between the surface boundary and the periodic boundary conditions
is always present in $d_0/a > O(1)$, even in the macroscopic (but ordinary) system size.


Fig.\ \ref{fig_cond}(b) shows  the magnetic-field dependence of
the conductivity for several potential ranges $d_0=a, 2a$ and  $3a$.
Again, we see a qualitative agreement in the numerics (solid curve)
and the analytics [Eq.\ (\ref{cond_zz_edge}),  dashed curve]
in long-range cases $d_0=2a$ and $3a$.
In the relatively short-range case, $d_0=a$, the conductivity does not follow 
the analytic curve but slowly approaching
the constant bulk conductivity Eq.\ (\ref{cond_zz_bulk}), indicated by the red dotted line.


\section{Discussion}
\label{sec_dis}

The negative magneto-resistance with the surface-mediated relaxation
is expected to be observed in the long-range disorder regime
where the Weyl node distance is much greater than the inverse of the potential range.
In actual Weyl semimetals recently discovered,
\cite{xu2015observation,weng2015weyl,lu2015experimental,xu2015discovery,lv2015experimental,huang2015weyl,xu2015experimental}
the typical Fermi-arc span is of the order of $\pi/a$ with $a$ being a few \AA,
so the long-range regime should be achieved when the potential length scale $d_0$
is larger than $\sim$1nm.
This condition would be satisfied when the screened charged scatterers dominate the system.
To observe the surface-mediated relaxation,
the linear to superlinear $B$-dependence  
and the anomalous thickness ($L_y$) dependence of the conductivity Eq.\ (\ref{cond_zz_edge})
would be direct evidences. 
In a weak magnetic field below the quantum limit,
the other Landau levels start to contribute to the transport,
and then the conductivity is expected to obey $B^2$ behavior
in the weak field regime. 
\cite{son2013chiral,burkov2014chiral,burkov2015negative}

While we assumed the magnetic field parallel to the surface in this work,
let us mention the effect of the field component perpendicular to the surface.
The recent theoretical studies showed that the $B$-field perpendicular to the surface
hybridizes the bulk Landau levels and the surface states.
 \cite{potter2014quantum,zhang2015quantum,baum2015current}
If the magnetic field is tilted from $z$ axis in our geometry, therefore,
we presume there is a mixing between the counter propagating bulk modes
via the surface, it would lead to an increase the electric resistance in $z$ direction.
In the experiment, the giant positive magnetoresistance under the magnetic field perpendicular
to the current was actually observed,
\cite{kim2013dirac,li2016negative,yang2015chiral,zhang2015detection,li2015giant,zhang2015observation,PhysRevX.5.031023,xiong2015evidence}
and it might be related to the field-induced surface-bulk hybridization.
We leave the detailed study of this problem for future work.




To conclude, we calculate the conductivity of the Weyl semimetal in 
${\bf E} \parallel {\bf B}$ geometry with a surface boundary.
We showed that the surface states always provide a relaxation path
between the two Weyl nodes,
and that definitely suppresses the diverging conductivity expected in the bulk model.
This offers an interesting example in which the surface boundary condition
and the periodically bound condition
result in completely different properties even in the macroscopic limit,
on the contrary to our intuition.

This work was supported by JSPS Grants-in-Aid for Scientific research (Grants No. 25107005).

{\it Note added}— The Fermi surface consisting of the surface states and the bulk Landau levels
has been also argued in very recent work.
\cite{PhysRevB.93.081103}

\appendix

\section{Analytical expressions for the conductivity}

We can derive an approximate analytical solution of the Boltzmann equation by assuming
$l_B \gg d_0$ and $\delta n_{\bk}$ is non-zero only in the bulk states.
Here, we show the derivation
of Eq.\ (\ref{cond_zz_bulk}) and (\ref{cond_zz_edge}).
In the above assumption, the wave function is well approximated by
the eigenstates of the zero-th Landau level in the continue Weyl Hamiltonian
\begin{align}
\psi(y)&=\begin{pmatrix}
             0 \\
             \phi(y-Y)
            \end{pmatrix}, \\
\phi(y)&=\left(\frac{1}{\sqrt{\pi}l_B}\right)^{1/2}\exp\left[-\frac{y^2}{2l_B^2}\right],
\end{align}
where the guiding center $Y=-l_B^2k_x$.
Using $\psi(y)$, we can derive an analytical formula for the matrix element
$\langle|V_{\bk\bk^\p}|^2\rangle$ as
\begin{align}
\langle|V_{\bk\bk^\p}|^2\rangle=&\frac{n_i u_0^2}{\sqrt{2\pi}L_xL_z\sqrt{l_B^2+d_0^2}}  \notag \\
                                                  &\times\exp\left[-\frac{(l_B^2+d_0^2)q_x^2+d_0^2q_z^2}{2}\right],
\end{align}
where $L_x, L_z$ are the system size for $x, z$ direction,
and $q_x=k_x-k_x^\prime$, $q_z=k_z-k_z^\prime$.
Substituting this into Eq.\ (\ref{def_tau}) and integrating on the Fermi surface,
we derive the relaxation time as
\begin{align}
\tau_s = \frac{2\pi\hbar^2v}{n_{\rm i}u_0^2} \frac{l_B^2+d_0^2}{1+\exp[-\pi^2d_0^2/(2a^2)]},
\label{app_tau}
\end{align}
and we see that $\tau_{\bm k}$ is independent of ${\bm k}$.

In the periodic boundary condition, $h_{\bm k}$ is independent of $k_x$
because of the uniform band structure along the $k_x$ axis, and the self-consistent
equation becomes
\begin{align}
h_+&=v+\frac{1}{1+\exp[-\pi^2d_0^2/(2a^2)]} \notag \\
                              &{~~~}\times(h_++\exp[-\pi^2d_0^2/(2a^2)]h_-), \\
h_-&=-v+\frac{1}{1+\exp[-\pi^2d_0^2/(2a^2)]} \notag \\
                              &{~~~}\times(\exp[-\pi^2d_0^2/(2a^2)]h_++h_-),
\end{align}
where $h_+=h(k_z=\pi/(2a))$, $h_-=h(k_z=-\pi/(2a))$.
Solving the above equations, we derive
\begin{align}
h_{\pm}&=\pm\frac{1}{2}\left(\exp[\pi^2d_0^2/(2a^2)]+1\right)v.
\label{app_h}
\end{align}
Substituting the derived $\tau_s$ and $h_{\pm}$ into Eq.\ (\ref{cond_zz}),
we get Eq.\ (\ref{cond_zz_bulk}).

In the surface boundary case, we completely neglect the intervally scattering,
and $\tau_{\bm k}$ is derived by omitting the exponential factor in Eq.\ (\ref{app_tau})
and written as
\begin{align}
\tau_l = \frac{2\pi\hbar^2v}{n_{\rm i}u_0^2} (l_B^2+d_0^2).
\label{app_tau_l}
\end{align}
In this case, we can solve the self consistent equation within each branch
and it becomes
\begin{align}
h_{\pm}(k_x)=\pm v+&\sqrt{\frac{l_B^2+d_0^2}{2\pi}}\int_{-L_y/(2l_B^2)}^{L_y/(2l_B^2)} dk_x^\prime h(k_x^\prime) \notag \\
                       &\times\exp\left[-\frac{(l_B^2+d_0^2)(k_x-k_x^\prime)^2}{2}\right].
\end{align}
Here, we extend the integration range
from $[-L_y/(2l_B^2),L_y/(2l_B^2)]$ to $[-\infty,\infty]$.
This procedure is approximately valid because the condition
$L_y\gg l_B$ is satisfied in the quantum limit, and
we obtain analytic solutions
\begin{align}
h_{\pm}(k_x)&=\mp v \left(1+\frac{d_0^2}{l_B^2}\right)
                       \left(l_B^2k_x^2-\frac{L_y^2}{4l_B^2}\right),
\label{hk_ana}
\end{align}
where $\mp$ is for the negative and the positive velocity branches $(k_z \approx \mp \pi/(2a))$,
respectively, and the above solutions satisfy the boundary condition $h_{\pm}(k_x=\pm L_y/(2l_B^2))=0$.
Substituting the derived $\tau_l$ and $h_{\pm}(k_x)$ into Eq.\ (\ref{cond_zz}),
we obtain Eq.\ (\ref{cond_zz_edge}).

We plot the derived approximate solutions of $\tau_{\bm k}$ and $h_{\bm k}$
in the top and middle panels in Fig. \ref{fig_nk}.
Eq.\ (\ref{app_tau}) and (\ref{app_h}) are plotted in Fig. \ref{fig_nk}(a),
and Eq.\ (\ref{app_tau_l}) and (\ref{hk_ana}) are in Fig. \ref{fig_nk}(b).
The solutions reproduce the numerically derived results
in the bulk Landau levels.

\bibliography{weyl_magneto}

\end{document}